# The Emergence of Bologna and its Future Consequences
## Decentralization as Cohesion Catalyst in Guild Dominated Urban Networks[1]

Rainer E. Zimmermann[2],

IAG Philosophische Grundlagenprobleme,
FB 1, Universitaet, Nora-Platiel-Str.1, D – 34127 Kassel /
Clare Hall, UK – Cambridge CB3 9AL[3] /
Lehrgebiet Philosophie, FB 13 AW, Fachhochschule,
Lothstr. 34, D – 80335 Muenchen[4]
e-mail: pd00108@mail.lrz-muenchen.de

Anna Soci

Dipartimento di Scienze Economiche,
Facoltà di Science Politiche,
Università degli Studi,
Strada Maggiore 45, I – 40125 Bologna /
Clare Hall, UK – Cambridge CB3 9AL
e-mail: soci@spbo.unibo.it

## Abstract

The following paper is on the emergence of observable complexity in urban networks visualized as product of essentially non-observable social processes. The methodology unfolded here draws on recent insight of econophysics in the strict sense under a top-down perspective of laying the foundations for a modern view to the evolution of dynamical structures in nature. The conception presented here deals with a section of ongoing cooperative research work being undertaken by the authors in collaboration with Giorgio Colacchio (now U Lecce). A first perspective as to the basic aspects of approach has been given in a joint paper in order to lay down the main ideas in some detail.[5] In the meantime, one of the authors (R.E.Z.) has had occasion to discuss the underlying concepts of this approach in a number of talks delivered within the years 2001 and 2002, notably to a study group in Berlin consisting chiefly of architects and city plan-ners.[6] Some material has been summarized in a recent talk presented to the German Physical Society.[7] This approach has also been topic of research during this same author's stay at the Institute of Advanced Studies of Bologna university while being a Senior Visiting Fellow there. On the other hand, all of this ongoing project is in turn itself a section of a wider research perspective which has been laid down earlier by this author in a number of papers and which comprises of a truly interdisciplinary discourse including philosophical as well as sociological, economic, and mathematical elements.[8] The cooperation mentioned here is thus also cross-related to the INTAS cooperation project „Human Strategies in Complexity" of which this author's Kassel research group is a member.[9] In the following, the exposition is solely focused on the project alluded to in the title however, rather than dealing with the wider framework in all of its interdisciplinary scope. Nevertheless, in the motivating introductory section the general context will be described shortly.

---

[1] Revised and extended version of a talk delivered at the 2003 EAEPE conference on „The Information Society. Understanding its Institutions Interdisciplinarily." in Maastricht. Cf. Collection of Extended Abstracts, 181-182.
[2] At the time of research on leave of absence as Senior Visiting Fellow to the Istituto di Studi Avanzati, Villa Gandolfi Pallavicini, Via Martelli 22/24, I – 40138 Bologna.
[3] Permanent Addresses.
[4] Present Address.
[5] R.E.Zimmermann, A.Soci, G.Colacchio (2001): Re-constructing Bologna. The City as an Emergent Computational System. An Interdisciplinary Study in the Complexity of Urban Structures. Part I: Basic Ideas & Fundamental Concepts. http://www.arXiv.org/pdf/nlin.AO/0109025 v2.

## 1. Introduction

Since the first seminal work on city dynamics presented by Friedrich Engels in his book on „The Condition of the Working Class in England" it has been demonstrated that essentially, the city is what Steven Johnson recently called „a pattern amplifying machine."[10] The city can thus be visualized as an orientation programming *matrix* which acts as a virtual interface facilitating the information processing performed by humans in their daily life. Walter Benjamin's unfinished work on Passages/Arcades can be found on the same line of argument. Johnson discovers more of the Santa Fe institute in Benjamin than of the Frankfurt school. In fact, the self-organized dynamics of an information processing network structure, which essentially *is* the city, is so much at the basis of nonlinear processes that it actually falls under the general framework of Alan Tu-ring's revolutionary paper on Morphogenesis.[11] The city we mean here is in fact a European invention[12]: Amazingly enough, its basic structure has remained almost invariant over the last thousand years, taking its origin from northern Italy, where by means of what we may call a *phase transition* (Johnson again) cities emerged due to a critical increase of production and energy flow mainly caused by innovative technologies (and thus economic consequences) in the agricultural sector. Bologna has been one of them at the time being founded not so much as a city in principle (as it is a Roman foundation), but because of the foundation of the *comune*.[13] The important point is that cities like Bologna can be discussed as a progressive (indeed *computationally self-processed*) mediating (urban) structure which is produced on the (non-observable) *micro-level* of social interactions, but which shows up in phenomenological terms on the (observable) *macro-level* of urban shape. Hence, what we do in this paper is to discuss the characteristic form of urban space in terms of these two levels of interaction, the macro-level being visualized here as emergent with respect to the micro-level. We discuss the former by means of *generative shape theory* in the sense of Michael Leyton[14], and the latter by means of *social network theory*.[15] Assuming that the explicit evolution of a city is a mixture of universal components and (context dependent) individual components, we basically ask: In the case of Bologna, what is universal, and what is individual? And also: Taking the case of Bologna as a paradigmatic example, what are the possible ethical implications of the results, and what can we learn from them for the future development of the city? And finally: What is the relationship of these conclusions to the urban planning program as presented by Bernhard Winkler to the city government in 1988/89?

---

[6] One of us (R.E.Z.) thanks here for co-operation Wolfgang Watzlaf, chairman of the Neue Akademie Berlin e.V. For the group's activities see www.neue-akademie-berlin.de .

[7] R.E.Zimmermann (2002): Decentralization as Organizing Principle of Emergent Urban Structures. In: AKSOE 14 (Urbane Systeme und Verkehrsdynamik II), Fachsitzungen des Arbeitskreises „Physik sozio-oekonomischer Systeme", Fruehjahrstagung der Deutschen Physikalischen Gesellschaft, Regensburg. http://www.arXiv.org/pdf/nlin.AO/0203012.
In the meantime, this paper has been published in a somewhat revised version in V. Arshinov, C. Fuchs (eds.), Causality, Emergence, Self-Organization, NIA-Priroda, Moscow, 2003, 36-55.

[8] R.E.Zimmermann (2000): Loops and Knots as Topoi of Substance. Spinoza Revisited. Long Version in: http://www.arXiv.org/pdf/gr-qc/0004077 v2. Short version under the title: Spinoza in Context. A Holistic Approach in Modern Terms. In: E.Martikainen (ed.), Infinity, Causality, and Determinism, Cosmological Enterprises and their Preconditons, Finish Academy of Sciences Colloquium (Helsinki), Lang, Frankfurt a.M. etc., 165-186 (2002). See also for an extended framework of discussion R.E.Zimmermann: Recent Conceptual Consequences of Loop Quantum Gravity, in three parts: http://www.arXiv.org/pdf/physics/0107061, http://www.arXiv.org/pdf/physics/0107081, http://www.arXiv.org/pdf/physics/0108026.

[9] Cf. the author's web pages http://www.uni-kassel.de/philosophie/iagphil/forschung/iagforsch4_5.htm as well as http://h2hobel.phl.univie.ac.at/asp/zimmermann.htm. The INTAS cooperation web page can be found under: http://www.self-organization.org/partners/kassel/zimmermann.html.

[10] Steven Johnson: Emergence, London: Penguin, 2001, 40.

---

[11] Alan Turing: The Chemical Basis of Morphogenesis. Phil. Trans. Roy. Soc. B (London), 237, 37-72, 1952.

[12] However, we learnt from Jack Goody and Roy Macleod that there is a mainstream belief in anthropology and related fields which would not visualize this sort of city as an European invention but instead as a universal institution which can be found practically everywhere on this planet. Contrary to that we still maintain the view that the medieval cities of Northern Italy comprise of a number of characteristics which are indeed of European origin only, especially with a view to the aspect of decentralized organization. This topic will be discussed in more detail in the forthcoming book of R.E.Zimmermann: *The Historical Centre of Bologna*. A Study in the Foundations of Emergent Urban Structures, the Unfolding of Social Memory, and the Form of Space. In preparation, 2004.

[13] Johnson places the actual emergence of these cities into Florence when the „guilds formed from the breaking apart of the Societas Mercatorum in the last decades of the 12<sup>th</sup> century." (Op. cit., 101) The *comune di Bologna* however has been formed already in 1117, and at that time the guild structure had been unfolded already for quite a while. See e.g. Salvatore Muzzi: Annali della Città di Bologna. Dalla sua origine al 1796. 8 vols., Bologna, 1840. Cf. I, 36-53.

[14] Michael Leyton: A Generative Theory of Shape. Berlin etc.: Springer, 2001.

[15] Although social network theory is a vast field for quoting relevant works, we base our approach mainly on the works of John Paul Boyd et al., Linton C. Freeman, Douglas R. White et al., and Frank Harary, especially in so far as they centre their approaches around theories of group actions which will play an important role as to their mediation with generative shape theory which is also essentially a theory of group actions.

## 2. The Macro-Level of Form: Urban Systems

The basic idea – as far as urban structures are concerned – goes back to work on problems of theoretical architecture by Hillier and Hanson[16] leading forward to a concept of „social logic" which is hold reponsible for the actual constitution of observable urban structures. In other words: Instead of being due to a completely random process of structure formation, the outcome of urban patterns within a city can be attributed to an *intrinsic logic* which is produced by the formation process itself, depending on internal and external conditions underlying the evolution of the whole urban system. Hence, what we can observe as an urban structure is the *emergent outcome* of the actual production process of the resulting urban space. This observable space (observable in terms of achitectural forms, traffic, communication, business and so forth) can be visualized therefore as the *macroscopic expression* of a network of *microscopic social interactions* among the agents (persons) who are the actual producers of that space.[17] This situation is similar to that of modern quantum physics which visualizes observable processes in everyday life as the emergent macroscopic outcome of underlying microscopic interactions and their superpositions. This is what in physics is called the concept of *decoherence*: Humans are „classical observers", i.e. they perceive and model the world in terms of classical physics dealing with the essentially Newtonian entities of space, time, and matter (because they are classical „entities" themselves). Hence, *the world is not as we perceive and model it*, but while modeling it, humans follow an algorithmic strategy which is based on their own intrinsic logic. The latter in turn is coupled to the world *as it is* (but as it cannot be perceived in principle), because humans (and their chief activity: reflecting) are themselves part of this very world. The conclusion of this is that in order to create models which are maximally self-consistent and plausible, one should look for models which can be embedded into the general framework of this „onto-epistemic" foundation. This fundamentally strategic viewpoint is one which guarantees *the genericity of the methodological procedure* being actually applied and thus *the universality of the underlying foundations*. It also puts forward a strict monism which couples physics through all possible fields of daily life up to politics. Hence, it also argues in favour of a consistent rationality of choice and by doing so selects explicit criteria for a practical concept of *ethics*.[18] The idea is then to carry over some of the conceptual aspects of physics to other fields of science. *Hence, we are talking about* econo-physics *proper here, in a strict sense, and we pursue in terms of a straightforward* top-down *approach rather than a bottom-up approach which is much more common in the field of*

---

[16] B.Hillier, J.Hanson (1984): The Social Logic of Space, Cambridge University Press. – See also B.Hillier (1996): Space is the machine. Cambridge University Press.
[17] Cf. H.Lefebvre (1991/1974): The Production of Space, Blackwell, Oxford.
[18] See details in R.E.Zimmermann (2001): Signaturen. NaturZeichen & DenkZettel. Zur morphischen Sprache der politischen Oekonomie. To be published. See also: *Graphismus & Repraesentation*. Zu einer poetischen Logik von Raum und Zeit. Magenta, Munich, 2004. And: Systematics as Normative Holism. Futura (Bull. Finish Academy of Future Studies), 3/2003, 55-69.

*urban studies.* This does not mean that mathematical approaches in particular are rendered to working metaphors which might eventually produce a kind of formal poetry. Instead, concepts carried over have to be re-interpreted quantitatively and qualitatively such as to become adequate for applications in the chosen field (which is not equal to the physical field, but conceptually a part of it). So what we gain in the end is a monistic viewpoint towards the totality of the world based on an implicit philosophical choice. But it is this choice in fact which secures the applicability and practical relevance of the produced results, in the first place.

## 2.1 A Combinatorial Approach

We shortly summarize here the essential aspects of Hillier's ideas[19] for a combinatorial approach to the layout of urban form. To this purpose, we reproduce a section from our first, fundamental paper[20] in order to remind on the explicit procedures being involved. The results of a first survey of actually performing these procedures will be only summarized in qualitative terms, because the detailed results will be topic of the second joint paper which is in preparation.[21]

*Motivation.* Urban layouts are being visualized in terms of more or less deformed grid shapes. The deformation from a regular grid can show up as either *axial deformation* (in that lines of sight and access are blocked by surfaces of building complexes) or as *convex deformation* (in that surfaces vary in their shapes all the time creating a number of patterns). Obviously, the intelligibility of a space will be related to the actual changes in the visibility field of a co-moving observer. This field in turn is determined by convex and axial aspects. The stronger the former, the larger the field, the stronger the latter, the smaller the field. In fact, there will be some kind of locally organizing centre of usually overlapping convex spaces which are called the *integration core* of the settlement. Now, we have then the following list of suitable measures:

*convex articulation* (of space) := number of convex spaces/number of buildings
The lower the values the more synchrony of the space is being achieved.

*grid convexity* := $(\sqrt{I} + 1)^2/C$, where I is the number of islands (blocks of continuously connected buildings), and C is the number of convex spaces.
The lower the values the stronger the deformation from a regular grid pattern.

*axial articulation* := number of axial lines/number of buildings
The lower the values the higher the degree of axiality.

*axial integration* (of convex spaces) := number of axial lines/number of convex spaces
The lower the values the higher the degree of axial integration.

---

[19] Cf. note 15 above.
[20] Cf. note 5 above.
[21] A. Soci, G. Colacchio, R.E.Zimmermann: Reconstructing Bologna. Part II. Forthcoming (2004).

*grid axiality* := $(2\sqrt{I} + 2)/L$, where L is the number of axial lines.
The lower the values the higher the degree of axial deformation.

*convex ringiness* := $I/(2C - 5)$
This measures the number of loops within urban space as compared to the maximum number of possible loops. Insofar it also measures the distributedness of the y-system which can be represented in terms of what is called y-map. This is a map in which convex spaces are being mapped as small circles, together with lines which signify their permeable adjacencies. Practically, a y-map transforms a convex map into a graph. The convex map is the least set of fattest spaces that covers the system. In fact, these are the instruments of visualizing the representation of a space as a set of syntactic relations, both of buildings and of other spaces. The synchrony of a space is then the quantity of space invested in these relations.

*axial ringiness* := $(2L - 5)/I$
This is the equivalent of the aforementioned with respect to axiality.

*axial connectivity* := number of lines a given line intersects
This is a self-explanatory measure of the connected integration of axial lines.

*ring connectivity* := number of rings a given line forms part of
This is the equivalent of the aforementioned with respect to loops.

From these combinatorial measures it is possible to derive a number of map variants explicating various details of the urban space:

*permeability map*
This is essentially a combination of convex spaces with buildings or bounded spaces, together with their connecting lines in terms of adjacency and direct permeability. Obviously, this has decisive consequences for the flow of communication in the urban space.

*decomposition map*
This is a variant of the aforementioned, insofar lines are indicated which link convex spaces constituted by front doors.

To both of the last two maps, converse maps can be drawn to illustrate the absence of flow lines of communication, i.e. locations of isolation.

*justified map*
This map gives the depth of a structure by connecting hierarchically accessible points to a chosen base point. Notably, it is important to analyze the clustering of spaces of equal depth, and the graph patterns of connecting lines.

*relative asymmetry* := $2(MD - 1)/(k - 2)$, where MD is the mean depth, and k the number of spaces considered.

The results of an adequate inspection of these characteristic numbers for the historical centre of Bologna can be summarized as follows: We find for Bologna a high synchrony, a low degree of grid deformation, and high degrees of axiality together with axial integration, and convex ringiness, pointing to a regular grid pattern. The *natural movements* (as determined by the grid structure itself rather than by the presence of specific attractors) are thus what we can call *de-labyrinthenizing*. The *portici* add to the closed framed space structure. Hence, this urban morphology leads to a close mediation of public and domestic spaces and thereby increases communicatibility.

### 2.2 Generative Shape

The basic idea of Leyton's approach is that geometric form is nothing static which would be „imprinted" into the human means of perception while learning the fundamental forms of environment, but instead something inherently dynamical in the sense that it is being *generated* by perception itself. In other words: (Cognitive) Perception is essentially (re-) construction. This idea of perceived shapes which are generative in the sense that they are permanently re-constructed in a dynamical way, leads to a straightforward formalism of group actions which can be taken as basic generators of form. A square e.g., is generated by means of a sequence of tracing a line element from the left to the right, say, and continue this process after a turning of ninety degrees. Hence, a square can be visualized as a product of translations and rotations such that

$$Z_2 \text{ w } R \text{ w } Z_4,$$

where the first group is the „occupancy group" giving the segmentation points of the actual quadratic form (cutting it out of the abstract network of possible squares), the second is the group of translations, and the third is the group of rotations by ninety degrees (having thus four elements). Note that the product operation w does not refer to the usual group product which acts upon normal subgroups, because normality would render the dynamics of generation impossible. Therefore, this product (called *wreath product*) acts upon non-normal subgroups.[22] This actually means that strictly, *there are no objects*, but only relational process structures. As form is generated all the time, each shape represents a history of deformations such that the shape in general can be thought of as a *memory* (of the processual sequences of symmetry breaking involved in generating the form). Recognition of form means then re-construction of its becoming. Both maximal transfer and maximal recoverability serve as boundary conditions

---

[22] A group N is normal subgroup of another group G, if for all g of G: gN = Ng, or equivalently, $gNg^{-1}$, i.e. conjugation which acts as an automorphism is being fulfilled. This being the case, the invariance of form is established while in the other case dynamical generation comes into play.

for the optimal selection of forms. The history of a form can then be visualized as the *unfolding of standard primitives* which are deformed in ongoing sequences. Standard primitives can be described in terms of iso-regular groups meaning that the group is decomposable as a control-nested structure (i.e. as a history), that each level is one-dimensional (i.e. cyclic or 1-parameter group according to whether it is discrete or continuous), and that each level is represented as an isometry. The following two tables give a short summary of properties of the wreath product and some of the standard primitives:

*Table 1: Wreath Product*

**Splitting Extension (Nesting)**

**$[G(F)_1 \times ... \times G(F)_n] \otimes(s) [G(C)]$**

**(The right-hand-side acts onto the indices of the left-hand-side which is the *normal subgroup* of the extension)**

**Normal subgroup = Kernel of a homomorphism**
**(A group X is normal subgroup of a group Y, if its conjugation by any member of Y acts as an automorphism.)**

$\Rightarrow$   **There is an automorphic representation:**

**T: $G(C) \rightarrow$ Aut { $[G(F)_1 \times ... \times G(F)_n]$ }**

**which is the transfer representation.**

**The semi-direct product above is the wreath product:**

**$G(F) \otimes(w) G(C)$**

Note that we have re-written the product symbols here for incorporating both the direct and the semi-direct (wreath) product. For the latter, $\otimes(w) = w$.

*Table 2: Standard Primitives*

**level-continuous**

| | |
|---|---|
| plane | $\mathbf{R} \otimes(w) \mathbf{R}$ |
| sphere | $SO(2) \otimes(w) SO(2)$ |
| circular cylinder | $SO(2) \otimes(w) \mathbf{R}$ |
| circular cone | $\mathbf{R} \otimes(w) SO(2)$ |
| torus | $SO(2) \otimes(w) SO(2)$ |

**level-discrete**

| | |
|---|---|
| cross-section block | $\mathbf{R} \otimes(w) \mathbf{Z}_4 \otimes(w) \mathbf{R}$ |
| planar face block | $\mathbf{R} \otimes(w) \mathbf{R} \otimes(w) \mathbf{Z}_4$ |
| cube | $\mathbf{R} \otimes(w) \mathbf{R} \otimes(w) \mathbf{Z}_2 \otimes(w) \mathbf{Z}_3$ |

In order now to apply this formalism of unfolding the shape history of a form to macroscopic problems in urban networks, we have to introduce the concept of *architectural mass groups* explained in detail by Leyton[23]: The idea is to compose architectural forms out of sequential deformations of standard primitives. In fact, applications in the architectural domain are comparatively easy, because forms can be *designed* here in the first place, and will take therefore a straight line of development starting with such primitives from the beginning on. In the following table the basic principles of architectural mass groups (i.e. group actions generating the deformations of a history of observable form within the urban environment) are listed:

---

[23] We follow here the description in his work, cf. note 14 above.

*Table 3:  Architectural Mass Groups*

**Principles**

**1.  A *mass group* is the unfolding of an alignment kernel.**

**2.  The alignment kernels are themselves unfolded hierarchically from a *base alignment kernel* up through subsidiary alignment kernels.**

**3.  A single symmetry group gives the entire unfolding structure.**

**4.  This maximizes transfer and recoverability.**

**The alignment kernel is the direct product of a number of clones of each *primitive*.**

**It basically contains three primitives: cube C, cylinder Cy, and sphere S. Hence:**

$$\{ [G]_S \times [G]_{Cy} \times [G]_C \}_U$$

**The number of clones will be determined by the wreath actions with the control groups.**

These principles can be applied then to single structures of buildings in a straightforward manner. Note that most of the structure details described here show up in numerous facade arrangements of historical cities of the Bologna-style and inherent layout organization. Indeed, in such cities, the individual form of a building is rendered almost unimportant as compared with the global outset, a point which is not only stressed in the university Rector's anniversary speech[24], but also mentioned in standard works on European cities.[25] Hence, the explicit structuring of places and squares (and of individual buildings as to that) vanishes in the background of a structural concentration on the large-scale layout of street nets, in the case of Bologna particularly framed by means of the *portici* dominating the historical centre.

---

[24] www.unibo.it (Storia dell'Università)
[25] Cf. R. Bartlett: The Making of Europe. Princeton University Press, 1993.

*Table 4:* *Unfolding the Alignment Kernels*

**1. Rectangular Masses 1, 2**

**{ ... } ⊗(w) { AGL (3, R) x { [ AGL(3, R) x e] ⊗(w) AGL(3, R) } x e } ⊗(w) AGL(3, R) x e**

**2. Cylindrical Mass 5**

**3. Tower mass grouping 6, 7**

**4. Lobby entrances 8, 9**

**5.   Floorplate Slicing**

**[K] ⊗(w) [G] ⊗(w) [G$_{slicing}$]**

**6.   Unfolding Space Volumes (Slicing)**

**R ⊗(w) Z$_4$ ⊗(w) R**

**7.   Column Grids**

**[SO(2) ⊗(w) R] ⊗(w) [R ⊗(w) R] ⊗(w) [Z ⊗(w) Z]**
     **column**        **deviation**        **grid**

**8.   Roof**

**ALG(3, R) ⊗(w) Z$_2$**

Hence, the *design process* altogether shows up as a *sequence of unfoldings*, and this in turn establishes a progressive process of symmetry breaking. In order to conclude we can thus say the following:

The evolution of shape (within this picture) is essentially a transfer structure which can be recovered by means of re-constructing the history of deformations of the isoregular groups being chosen for characterizing the standard primitives in the first place. That re-construction follows an explicit *process grammar* which tends to achieve a maximal recoverability of the likewise maximal transfer by means of uncovering a sequence of symmetry-breaking phase transitions. Because of the equivalent wreath product structure of the human perceptual and motor systems, this process grammar serves as the basis for a generalized *morphic* language which is at the root of social communication. Hence, the degree to which maximality of recoverability and transfer is being achieved determines the maximality of communicability of the forms in question. This is the nucleus of the assumed isomorphism between the observable macroscopic (architectural) form of urban structures on the one hand and the (essentially non-observable) microscopic form of the social networks which are actually producing the former such that it is emergent with respect to the latter.

### 3. The Micro-Level of Form: Social Networks

In physics, the starting point is the discrete combinatorial structure of space (and time) which is representing a more fundamental level of modeling the world as compared to the classical level of modeling the world as it is actually perceived in terms of a smooth space-time continuum. Hence, space (and matter occupying space) is visualized as a sort of „average" taken over that fundamental level. In other words: An explicit *averaging procedure* is looked for which can generate a plausible approximation of observed space so as to eventually *reconstruct the emergence* of the classical macroscopic world from underlying microscopic interactions. It is the theory of loop quantum gravity which offers a procedure for actually doing so: Bombelli introduces (gravitational) loop states that should be a good approximation to classical geometry. [Take a Riemannian manifold space, sprinkle points on it randomly and form the corresponding Voronoi diagram, label the edges with spin numbers, then average over all possible ways of randomly sprinkling these points to finally obtain the *random weave state* which approximates the geometry we started with in the first place.[26]] In fact, as it turns out, this procedure is a direct application of the concept of *spin networks*, originally introduced by Penrose, and later modified by Smolin and Rovelli: These networks represent the spin interactions of abstract particles by purely combinatorial means, because the spin numbers labeling the edges of the respective graph are essentially irreducible representations of the groups relevant for this sort of physics, in particular of the type SU(2) and SL(2, C).

---

[26] L.Bombelli (2001): Statistical Geometry of Random Weave States. http://www.arXiv.org/pdf/gr-qc/0101080.

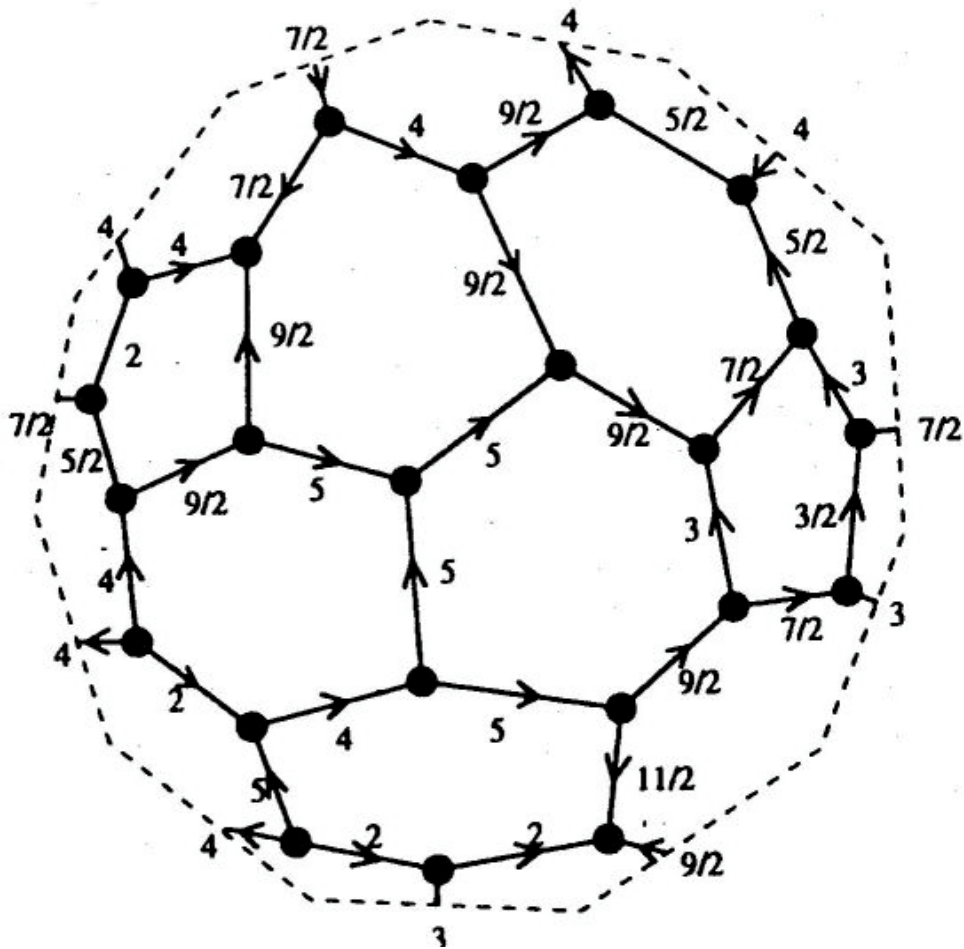

Hence, in the figure displayed above, the numbers indicate the matrix entries of the representation of group SL(2, C). And we can interpret the network as one of such interactions that transport information expressed in terms of these numbers. In other words: While *processing* the numbers (or equivalently, the information) the above network *represents* space, provided it is being observed by human (macroscopic) observers. Hence, the network structure displayed above can be visualized as a model of a computational process.

It is thus the latter group SL(2, C) that secures the interpretation of spin networks in terms of fundamental (quantum) computers performing an *intrinsic process algorithm* by which space and time (and matter) are actually being „produced" permanently. Hence, a change on the spin network level corresponds to some observable motion on the macroscopic level of the classical world. The important point is that in order to express these microscopic motions one needs combinatorial procedures only: The approximating average is produced simply by taking various sums and products over the amplitudes of respective edges and vertices of the diagrams. In order to include temporal sucession in the picture, *spin foams* have been introduced which are simplices in the triangulations of some suitably chosen space. Essentially, spin foams constitute *updating steps* in the above network, i.e. a sequence of such networks labelled with different numbers representing different states. The average is then expressed by the *partition function* which is the sum over all possible amplitudes. [In fact, spin foams are generalized Feynman diagrams within the group field theory approach: Feynman diagrams in quantum field theory compute the expectation value of some observable measured in the future, given information about the past. Spin foams do essentially the same.]

In utilizing the results of loop quantum gravity one can establish elegant relationships among most recent fields of physics, mathematics, and computer science, as there are *loops* and *knots* as well as *categories* on the one hand, and *cellular automata* on the other. All of these have contributed to the recent understanding of emergent and evolutionary processes.[27] As has been shown by Hillier and Hanson, and as has been discussed in earlier work mentioned above[28], this same procedure can be conceptually utilized in order to approximate the geometry of structures in urban space. One of the simplest viewpoints is here to visualize a city as a network of paths: Buch and Schubert have discussed such a model, although under the perspective of simulating the fitness optimization of essentially trivalent graphs only.[29] Another very interesting model which operates very much in the vicinity of our own model is that of Luca Caneparo et al. partially based on the idea of utilizing cellular automata.[30]

Note that the macroscopic counterpart of form related to the microscopic situation as it is described above can be visualized as a similar process of computation as it is being perceived by the human observers and actually reflects what is going on at the non-observable level such that an *epistemic parallelism* is established which enables inference from one to the other level. In fact, this macroscopic equivalent is a *cobordism* mapping which reproduces the dynamical change of spatial form as displayed in the next figure.

---

[27] I have discussed some of these aspects in more detail in: Classicity from Entangled Ensemble States of Knotted Spin Networks. A Conceptual Approach. http://www.arXiv.org/pdf/gr-qc/0007024. For cellular automata see also A.Schatten (1999): http://www.ifs.tuwien.ac.at/~aschatt/info/ca/ca.html.

[28] See notes 5 and 7 above.

[29] S.Buch, A.Schubert (1998/99): http://plato.igp.uni-stuttgart.de/~sbe/wwwork/text/thema8.html.

[30] See e.g. L. Caneparo, M. Robiglio: Evolutionary Automata for Suburban Form Simulation. CAAD Futures 2001. Reprint from the website, Dip. Prog. Arch., Politecnico di Torino. The important point here is that this research project also focuses on the qualitative analysis of urban form.

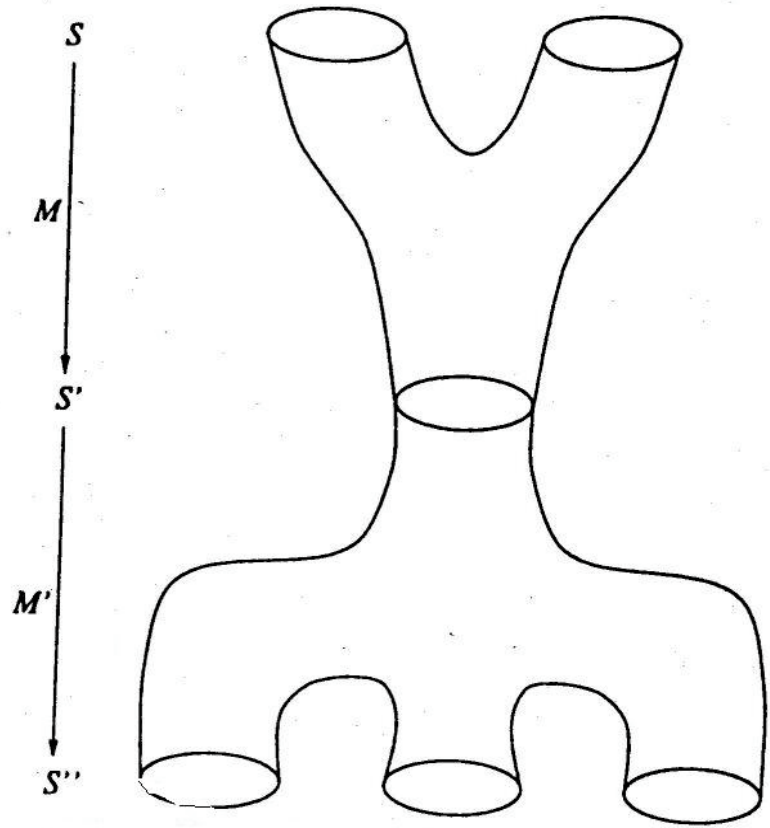

Here, a composition sequence of cobordisms describes the change of form as displayed in terms of three-dimensional spaces S labelled by a four-dimensional space-time manifold M. (Hence, time as „medium of change" is being absorbed into the geometry proper.) Emergence then (in the sense of physical decoherence) can be expressed by means of a commutative diagram of the form

$$\begin{array}{ccc} \text{nCob} & \rightarrow & \text{Hilb} \\ \uparrow & & \Uparrow \\ \text{SpinF} & \rightarrow & \text{Hilb} \end{array}$$

where SpinF is the category of spin foams, nCob the category of n-cobordisms, and Hilb the category of Hilbert spaces. The mappings are the appropriate functors (and thus *theories* at the same time). This principle of commutative diagrams remains invariant independent of the field of applications chosen. Hence, there is dual „column" to the left-hand-side of the above diagram, of the form

$$\begin{array}{cc} \leftarrow & \text{NEG} \\ & \uparrow \\ \leftarrow & \text{SpinF*,} \end{array}$$

such that it completes the diagram as its right-hand-side.

We collect then the starting points again: The city is visualized here as something which is being produced by a kind of programme satisfying its own intrinsic logic. The city can be interpreted therefore, as a computational process following its own prescribed algorithm. This process however can only be observed by means of decoherent structures emerging from this computational process on the classical level of human perception: The urban world is not as we observe it. Hence, there is a dynamical mediation between a *geometry of discourse* forming abstract spaces of communication produced by the ongoing social interactions on the one hand, and a *geometry of observable urban space* on the other. So the social interactions actually *produce* urban space, and it is how this space is being observed what constitutes the materialization of the average communication representing the outcome of the underlying interactions.

It is important to notice here that „complexity", „emergence", and similar concepts describing urban structures with a view to general evolutionary principles, are always *epistemic concepts* rather than properties of the world. (Contrary to the latter they are properties of humans modeling and thinking.)[31] So what we are doing all the time is to apply models to models (to our *modal* world, not to the *real* world), but this is not only a practical advantage (because of the resulting universality of certain methodological procedures so that we can profit from conceptual analogies between physics and politics, so to speak), but also something which is at the same time part of he (real) process itself and securing therefore that we nevertheless speak about the world *that there is* after all. (Hence, the onto-epistemic viewpoint of this very approach.) In order to actually translate forms in the „space of discourse" into forms in the „space of observed structures", we pursue according to the physical analogy. As mentioned above, this does not mean however that we would transform physics into its metaphorical counterpart – provided of course that we properly re-define the combinatorial entities we are actually utilizing in the process of translation. So the idea is to label interaction networks by numbers which characterize the strength and quality of interactions. Luckily, as it turns out, social interactions can be mapped to trivalent graphs, too (as is the case in physics). This is so, because (as Sartre has shown earlier in 1960) any bundle of social interactions can always be reduced to a bundle of pair interactions. In terms of graph theory, this can be expressed by edges of a trivalent graph.[32] The appropriate fundamental network structures of trivalent graphs of interaction can be reconstructed by utilizing the Voronoi procedure and checking the observable results on urban structures against the average geometry actually being produced. Criteria for what may be relevant in terms of observable structures have been offered by Hillier and Hanson. They have been discussed in the Bologna paper quoted earlier.[33]

---

[31] B.Edmonds (2001): What is Complexity? – The philosophy of complexity per se with application to some examples in evolution. Manchester. http://alphard.cpm.aca.mmu.ac.uk/. Cf. also: R.K.Standish (2001): On Complexity and Emergence. http://www.arXiv.org/pdf/nlin.AO/0101006. And: J.P.Crutchfield (1993): The Calculi of Emergence: Computation, Dynamics, and Induction. Santa Fe Institute, Research Paper SFI 94-03-016. Also in: Physica D (1994) special proceedings issue (Oji International Seminar „Complex Systems – from Complex Dynamics to Artificial Reality).

[32] The One interacting with the Other under the Look of the (constituting) Third is discussed in some detail for the first time in J.-P.Sartre: Critique of Dialectical Reason, Gallimard, Paris, 1960, in chapter I B.

Hence, in order to adequately treat social networks as underlying micro-structure of the observable urban macro-structure, we base our ideas on the described network equivalence: That is, vertices/actors (showing up as intertwining operators) and edges/relations among actors (showing up as group representations) determine the mapping of the social group dynamics actually constituting the processes in question. So essentially, problems of interactions in social networks can be studied in terms of *walks* as characterized by the appropriate *adjacency matrix* such that the matrix of n-th order gives the frequency of paths of length n which connect each actor with each other actor. Social interactions show up after all, as a a *percolation* problem mapping the *connectivity* of actors in the network. Characteristic measures of the network such as robustness (stability), combinatorial metrics (distance) for actors, in particular geodesic paths, power distribution and centralization versus decentralization phenomena constitute the complete set of problems which can be studied on arbitrary (and thus also on social) networks.

It is important to note that *group patches* (cliques) enhance the stability of such networks. Cliques and subgroups are nothing but maximally complete subgraphs of the network's representation: A vertex then is member of a clique of size n, if it has direct ties to n – k members of that clique. The maximal group of actors all of whom are connected to some number k of other members of the group is called k-core. The idea is also to determine the actual *lambda sets* of the group, i.e. the ranking of each of the relationships in terms of importance by evaluating how much of the flow among actors goes through each link. Identifying then sets of actors which, if disconnected, would most greatly disrupt the flow among actors, we find how the flow is actually related to the number of actors in the neighborhood leading to pathways.

Note finally that two vertices u and v of a labeled graph are automorphically equivalent, if all the vertices can be relabeled to form an isomorphic graph with the labels of u and v interchanged. In other words: Exchanging actors has not effect on the distances among all actors. Hence, clique data can also be visualized as a Galois lattice: This is a triple (A, C, M) in which A is a set of actors, C is a collection of cliques, and M is a binary relation in A x C. Then, (a, c) ∈ M means that actor a is a member of clique c.

In the case of Bologna, there is another important aspect of the city's form which is represented in terms of the ancient city gates (of the former modern city walls demolished mostly by now). Winkler has pointed to the significance of these gates when writing his book on space and mobility: He visualizes the left fragments of the former gates surrounded by places and squares which manifest the vital role of traffic flow over the boundaries of the historical centre as loci of a virtual unity between space and community.[34] As such the gates gain the explicit connotation of a *chronotopos* in the sense of Bachtin: „Everything lightlike and shadowlike of passed epochs is there engraved."[35] Hence, they focus the spatial experience of the city lived by the pedestrians passing and populating its streets.[36] This is in fact the idea of the sensitive city as discussed by Boesch: to leave spaces free (empty) in order to generate spaces of free play and creativity.[37] Ways to introduce concrete criteria for applying this idea show up, among other things, in the organization of quarters primarily accessible to pedestrians[38], and one method among many is the transformation of the concept of city gates.[39]

---

[33] See note 5.

What we notice when trying to put these aspects to a sound formalization within the framework developed here, is that as to city gates we can formulate a kind of „holographic principle", namely that *the structural characteristics of the city's historical centre can be mapped onto its city gates where, in their transformed manner, reflect themselves completely*. We will discuss this in more detail at another place.[40] For us here, it is noteworthy however that when applying this principle to the historical centre of Bologna, we indeed reproduce the structural layout of the city as presented earlier by Winkler by independent means of research.[41] For him, it was important to represent the city structure by the patterns prescribed by the gates and the lineage of connectivity achieved by main pathways for pedestrians. Implicitly, this uncovers the symbolical role the now transformed gates actually play (even if they are completely absent which is the case twice). For our purposes we can take this schedule as a „skeleton map" of the city displaying its main structural characteristics. The important point here to notice is that these results do not only tell us something about the traffic flow in the city and the input and output of economical functions. Instead we also gain information about the social structure of the groups actually inhabiting the city, and we learn details about the interaction between the symbolism of the gates and the fundamental social motions as performed by the pedestrians.

## 4. The Historical Foundations

It is quite obvious though that all the structural results obtained about the city of Bologna cannot be separated from the city's explicit historical development. Unfortunately, as it turned out, it is almost impossible to find historical material as to the founding of the city's *comune* around the year 1117. The point is that, not by coincidence, this founding of the *comune* is tightly connected with the founding of Europe's first university and the propagation of Roman law as first chief subject of this university's teaching emerging from the research undertaken at the time with a view to the problems posed by the Investiture Contest, and, by the way, putting forward positions of one of its protagonists, the emperor. But it appears to be extremely difficult to obtain sufficient knowledge about the proceedings during this critical epoch. Although it is not the appropriate place here to discuss the city's history in detail, we would like to point to a number of important points which are revelant with a view to the topic discussed. The *first point* concerns the origin of this „stronghold" of Roman law: As Radding has explained in detail, during the reign of the Lombard codes of law the Roman law had never been abandoned altogether. In particular, the Lombard law could not penetrate some areas such as that around Ravenna including Ferrara and Bologna.[42] An explicit period of „mixed jurisprudence" emerged and dominated until the important time of the Investiture Contest when problems of law became a vital part of the ongoing proceedings. Obviously, this influence cannot be separated from aspects of the rising commerce which is, especially for Bologna, a very significant ingredient for the understanding of the political structures of the time.[43] The *second point* concerns the origin of the university: The university's Rector gave an appropriate definition at the occasion of his speech for the 900[th] anniversary of the university's foundation. He defined a university as an institution based on the dual principle of independent scientific research and teaching.[44] Independent of whether this principle could have been secured through the centuries or not, the basic idea appears to us as something which would be untypical for similar institutions, e.g. in Asia: The strict rules of a feudal system as the Chinese one would prohibit independent and thus scientific research of that kind from the outset. In other words: The organizational structure of the European university is, contrary to (as far we can see) all similar institutions on other continents, metastably poised at the boundary between political influences as put forward by the Church on the one hand, and the Emperor on the other. This indeed *critical* position is the necessary (though not sufficient) condition for scientific research and teaching. And hence, we would agree to the statement that Bologna can count as having produced the first European university. The *third point* is that the complex interaction of a number of aspects during the critical period of founding the university obscures the real processes of formation (and appears to be ill-documented or at least not in the focus of historical research so far). In the city chronic of Salvatore Muzzi, the years between 1112 and 1117 seem of prime importance, and as it appears, the foundation of both the *comune* and the university are closely related to the political events within the ongoing contest between emperor and pope. Obviously, all fundamental achievments of Bologna such as the de-centralized quarter policy have been present already at this time.[45] But the question why Bologna would have been able to maintain its independence for a very long time amidst an ongoing political contest of most significant influences remains unanswered. This will be topic of further work to be published.[46]

---

[34] B. Winkler: Stadtraum und Mobilitaet, op.cit., 71, 186.
[35] Ibid., 43: „Tutte le luci e le ombre delle epoche andate vi sono scolpite."
[36] Ibid., 27 sqq., 56.
[37] H. Boesch: Die sinnliche Stadt. [The sensitive city] Essays zur modernen Urbanistik. Nagel und Kimche, Zuerich, 2001, 54.
[38] Ibid., 45, 66, 81.
[39] Ibid., 143.
[40] R.E.Zimmermann: The Holographic Principle of a City: The Case of Bologna. Forthcoming. – For the Bologna gates project see also the website: http://h2hobel..phl.univie.ac.at/asp/zimmermann.htm .
[41] B. Winkler: Stadtraum und Mobilitaet, op.cit., 205.

---

[42] C.M.Radding: The Origins of Medieval Jurisprudence. Pavia and Bologna 850-1150. Yale University Press, New Haven, London, 1988, 33.
[43] Ibid., 113 sq.
[44] www.unibo.it (Storia dell'Università)

## 5. Consequences

To actually choose the city of Bologna as an example for utilizing the methodological procedure described here has straightforward reasons indeed: As to the explicit construction of its urban form, Bologna has had a revolutionary past. In particular, with respect to its historical centre, the structural development of a progressive conservation of historical urban substance has been prepared in an exemplary (if not paradigmatic) manner. The original programme of sanitation and rehabilitation ( = re-construction) and its mediating presentation (Piano per il centro storico, 21st July 1969, cf. Pier Luigi Cervellati: Commune di Bologna, Centro storico, 1970), both the outcome of a period of scientific co-ordination by the study group of Leonardo Benevolo between 1962 and 1965, have been milestones of modern city planning. For the first time, as far as we can recognize today, the methodological conception of a scientific as well as historical and morphological urban analysis has been laid down with a view to treating the historical centre as a complete monument in its totality.[47] A number of innovative concepts have been decisive for the practical application of this approach in daily urban life for a long time. But first of all, it is the principle of an explicitly *de-centralized public organization* which has been the main pillar of this urban structural architecture: The idea was to take *quarter councils* and *quarter assemblies*, respectively, as means of a democratic integration of decision making by transparent, self-organizing groups as counterparts of the private sphere of the family organization (decentramento). An immediate corollary of this approach was the necessity of establishing an urban structure of *nearness and vicinity* aiming towards the concrete public life *in the place* rather than of drafting an abstract scheme of transitory traffic flows. As documented later at the first world conference of urban traffic (10-12 June 1974) when the „declaration of Bologna" was signed, *mobility* is visualized within the framework of this conception as one of the most essential necessities of human life.[48] Its fulfilment should be satisfied therefore such that security, comfort, convenience, and swiftness can be actually harmonized. In this sense, transport is being defined as a *public service* which must not be organized according to principles of profit, but only according to its social utility. In particular, the rights of the pedestrian should be visualized as part of the human (declaration of) rights.[49] In the sense of this declaration, the administration of Bologna had already introduced a *utility tariff* of public transport with a zero rate for all during rush hour traffic, and free of charge for pupils and students during school hours and university terms, and for old age pensioners altogether. The main underlying objectives of this approach are still today part of the urban planning objectives in so far as they outline the basic achievement of „a sustainable mobility through a decrease in energy consumption, a reduction of pollution, the offer of adequate accessibility [as well as] the recovery of urban areas unduly invaded by cars."[50] The detailed consequences of this conception have been discussed in detail in the already quoted paper.[51] For us here, it is important to realize that the city of Bologna can be treated as a paradigmatic example for all what we said above with regard to the evolutionary aspects of urban computational structures. This is not particularly so because for a long period of Bologna's city government, the urban conception would have been put forward according to theoretical principles of Marxist political economics. Much on the contrary, this is particularly so, because the urban conception has been introduced *as if* there were universal principles of algorithmic computation deriving from a fundamental logic of *anthropological* rather than merely political scope. And this is what renders the case of Bologna paradigmatic in this sense. Therefore, the very practical case of reconstructing the historical centre of this city with a view to the aforementioned intrinsic logic has been chosen as a prime example in order to explicate the ideas as they have been laid down here.

---

[45] S. Muzzi: Annali della Città di Bologna. Dalla sua origine al 1796. Op. cit., I, 48 sqq. – See also in Radding: The Origins, op.cit., 158-164 and W.Steffen: Die studentische Autonomie im mittelalterlichen Bologna, Lang., Bern etc., 1981, 34.

[46] R.E.Zimmermann: The Historical Centre of Bologna, op.cit.

[47] L.Jax (1989): Stadterneuerung in Bologna, 1956-1987. Zum Aufstieg und Niedergang der Quartiersdemokratie. [Urban Reconstruction in Bologna] Koeln: Kohlhammer.

[48] This is also the main concept of the fundamental work by B.Winkler: Stadtraum und Mobilitaet (Spazio urbano e mobilità), avedition, Stuttgart, 1998.

## 6. Conclusions

The results can be summarized as follows: *1. Macro-level. 1.1 Combinatorial Approach:* We start with applying the essentially combinatorial *morphic language* introduced by Bill Hillier et al. and find for Bologna a high synchrony, a low degree of grid deformation, and high degrees of axiality together with axial integration, and convex ringiness, pointing to a regular grid pattern. The *natural movements* (as determined by the grid structure itself rather than by the presence of specific attractors) are thus what we can call *de-labyrinthenizing*. The *portici* add to the closed framed space structure. Hence, this urban morphology leads to a close mediation of public and domestic spaces and thereby increases commu-nicatibility. *1.2 Generative Shape*: The essential structure here is the *wreath product* $\square$ of control group and fibre group such that the former sends the fibre group copies onto each other. The system of all the copies is a direct product while it is related to the control group by a semi-direct product. The respective fibre group will be called *alignment kernel*, and the idea is that the control group is then a hierarchical process of unfolding the maximally collapsed form which is this alignment kernel. This is basically a process of transfer uncovering the generative history of the actual shape observed and so utilizing in fact shapes as *memory of matter*. The important point is that the human perceptual and motor systems are both structured as n-fold wreath products of that sort. Hence, the unified view of shapes and human behaviour. For systems in architecture, the unfoldings of alignment kernels are called *mass groups*. So for buildings constituted of sufficiently complex deformed structures, the product structure of group actions is essentially of the form

$$K \ w \ G_m \ w \ G_s^{a...bP},$$

where K is the alignment kernel, $G_m$ is the unfolding group of the massing structure, and $G_s$ is the unfolding group of the slice structure. The unfolding of space volumes, the structural column grids of buildings and their deviation from symmetry group patterns are being included into this by means of further wreath products. For Bologna, we find that shape primitives can be recovered easily, in particular with a view to the iso-regular groups underlying visual perception. Symmetry axes play an important role here. Hence, we argue that in this case, the adaptation of new structures to old ones is facilitated, which increases the communicatibility of the urban system while the maximization of transfer and recoverability (what Leyton defines as criteria for insightful behaviour) is permanently perceived in cognitive terms and communicated to others. This is at the root of transforming reflexion into action. *2. Micro-level. Social Networks*: Utilizing the theory of semi-groups and Galois lattices, we construct a social network for the historical centre of Bologna by means of the snowball method beginning with focal sets of actors. We represent this network by signed digraphs and look for their connectivity properties. For Bologna, we find that connectivity is strongly amplified by „clique patch" (social) subgroups introducing an explicit *decentralization* into the urban network structure. This adds to the stability and robustness of the network's connectivity.[52] These results are compatible with those for the macro-level which is the emergent result of the interactions on the micro-level. *3. Historical Foundation*: The origin of *decentralized centre cities* whose neighborhoods within the metropolitan structure are themselves polycentric structures (Johnson) is due to the ancient *guild structure* of the town's social organization. Because originally, academic groups (of professors and their students) were visualized as nothing but one form of guilds among others[53], it is not a coincidence that the founding of the *comune* is closely tied to the founding of the world's first university, and that it is Roman law which plays an important role here. *4. Consequences*: It is argued that highly connected decentralized city structures as they are widespread in today's Europe have to be cherished giving some ethical implications of this with respect to the concept of solidarity. We remind on the 1974 Bologna declaration. It is found that Winkler's urban mobility plan of 1988/89 has been an important step into the right direction.

---

[49] M.Jaeggi, R.Mueller, S.Schmid (1976): Das rote Bologna. [Red Bologna] Zuerich: Verlagsgenossenschaft. Here: 103sq.
[50] Urban Traffic Regulation Plan, Bologna, Italy, http://www.eltis.org/data/67e.htm. Here: 1.
[51] See note 5.

---

[52] This also accounts of the clustering inertia of city businesses discussed in the Krugman-Schelling model. From the Santa Fe group especially James P. Crutchfield and Melanie Mitchell have discussed decentralization. See also Michael Resnick: Turtles, Termites, and Traffic Jams. Explorations in Massively Parallel Microworlds. Cambridge (Mass.): MIT Press, 1994.

## 7. Acknowledgments

One of us (R.E.Z.) would like to thank Roberto Scazzieri, the director of the Institute of Advanced Studies in Bologna, and the institute's committee, for their kind invitation to stay at the institute as a Senior Visiting Fellow from February until April 2003 within the peaceful and stimulating atmosphere of the beautiful Villa Gandolfi Pallavicini encircled by a large, mysterious park guarded heavily by two scholarly pheasants. For illuminating discussions he also thanks his co-fellows Jack Goody (Cambridge, UK) and Roy Macleod (Sydney). We would also like to thank Luca Caneparo (Torino) for his kind interest in our work and the exchange of information relating to the topic.

## 8. Selected References

---

[53] Jacques Le Goff: Les intellectuels au Moyen Age. Paris: du Seuil, 1984.